\documentclass[aps,pre,nofootinbib, onecolumn,showpacs,superscriptaddress,groupedaddress,rmp]{revtex4-1}
\usepackage{scrextend}
\usepackage{verbatim}
\usepackage{graphicx}
\usepackage{hyperref}
\usepackage{mathtools}% Loads amsmath
\usepackage{bm}
\usepackage{lipsum}
\allowdisplaybreaks
\usepackage{amsmath}

\usepackage{xcolor}
\usepackage[makeroom]{cancel}
\usepackage{subcaption}
\usepackage{titlesec}
\usepackage{amsmath}

\begin{document}
	\title{ Breakdown of additivity of transition rates in systems connected to multiple thermal reservoirs }
	 
 	\author{Vaibhav Wasnik}
 	\email{wasnik@iitgoa.ac.in}
	
 	\affiliation{Indian Institute of Technology, Goa}
	
	\begin{abstract}
		
In stochastic thermodynamics, it is commonly assumed that for a system coupled to multiple thermal reservoirs, the transition rates between two energy levels are additive across baths.
In this work, we first demonstrate through an explicit construction of two subsystems—that are parts of a single composite system, each coupled to a distinct thermal reservoir—that while each subsystem individually evolves Markovianly, their joint evolution is inherently non-Markovian. {We then present a general algebraic argument showing that, even if one assumes a Markovian description with additive transition rates, the steady-state condition imposed by the master equation leads to an inconsistency}. The analysis identifies the precise structural limitation that arises in describing systems simultaneously interacting with multiple baths within a Markovian framework.

	\end{abstract}
	
	\maketitle

 	 \section{ Introduction } 
 	 Traditionally  thermodynamics has  concerned itself with the behaviour of equilibrium and close to equilibrium    macroscopic systems. The microscopic origin of these  thermodynamic laws has been achieved using the tools of equilibrium statistical mechanics. 
 	 Over the past few decades  progress has been made towards extending the concepts of traditional thermodynamics to mesoscopic systems, which has led to understanding the far from equilibrium behavior of these systems \cite{stochastic1}-\cite{stochastic8}. The second law of thermodynamics  finds  itself extended to fluctuation theorems that are concerned with  individual trajectories traversed by the systems as they exchange energy with the surroundings.

 	 Majority of works in the field are concerned with systems that evolve in a Markovian fashion. The master equation for probabilistic evolution of such a system that can take energy values labelled by $\epsilon_j$ is given by  
 	 \begin{eqnarray}
		\frac{dP(\epsilon_i)}{dt} = \sum_{\epsilon_j} W_{\epsilon_i,\epsilon_j}P(\epsilon_j) - \sum_j  W_{\epsilon_j,\epsilon_i}P(\epsilon_i) 
		\label{master_equation}
 	 \end{eqnarray}
 where $P(\epsilon_i)$ is the probability of the system to be in a state of energy $\epsilon_i$ and $W_{\epsilon_i,\epsilon_j}$ is the rate at which transitions happen from state of energy $\epsilon_j$ to $\epsilon_i$.  For systems connected to a single thermal reservoir at temperature $T$, the transition rates obey detailed balance condition
  \begin{eqnarray}
 W_{\epsilon_i,\epsilon_j} e^{-\beta \epsilon_j} = W_{\epsilon_j,\epsilon_i} e^{-\beta \epsilon_i}
 \end{eqnarray}
 where $\beta = \frac{1}{k_B T}$.	Systems in contact with multiple thermal reservoirs are not only important theoretically where they evolve to non-equilibrium steady states, but are also important from a practical viewpoint since most applications such as engines, pumps etc involve systems connected to multiple reservoirs.  Many works  on the subject, \cite{stochastic1},  \cite{demon1},   \cite{demon2}, \cite{sum_1},  \cite{multiple1}, \cite{multiple2},  of    a system in contact with $n$ thermal reservoirs at temperatures $T_k$ with $k \in [1,n]$ make an assumption   that  
 	 \begin{eqnarray}
 W_{\epsilon_i,\epsilon_j} = \sum_{k=1,n} W^k_{\epsilon_i,\epsilon_j}
 \label{sum}
 	 \end{eqnarray}
    where $ W^k_{\epsilon_i,\epsilon_j}$ is the rate at which   transitions  happen between states of energy $\epsilon_j$ to $\epsilon_i$ if the system was only in contact with a reservoir at temperature $T_k$.	
    
\begin{figure}
	\centering
	\includegraphics[width=0.7\linewidth]{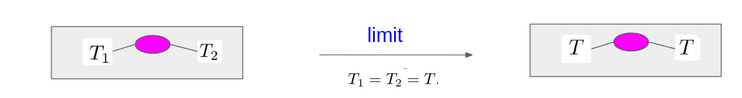}
	\caption{}
	\label{fig:fig11}
\end{figure}

%     The $W^k$'s satisfy the detailed balance condition
%    \begin{eqnarray}
%     W^k_{\epsilon_i,\epsilon_j} e^{-\beta^k \epsilon_j} = W^k_{\epsilon_j,\epsilon_i} e^{-\beta^k \epsilon_i}
%    \end{eqnarray}
% 	  
 	   Eq.\ref{sum} seems to work well as an equation because  it even to lead to the second law of thermodynamics \cite{stochastic1}.  However  is this  reason enough for validity of Eq.\ref{sum}?  
The way such a formula is justified is following. Imagine we have a system  as shown in Fig.\ref{fig:fig11}, where one part of the system is connected to a bath temperature $T_1$ and another to a bath at temperature $T_2$. If one assumes the system can have energy levels $\epsilon_i$ where $i \in [1,N]$, then in time $dt$ the   system can make a jump starting from energy level $\epsilon_i$ to $\epsilon_j$, if it exchanges    energy either with reservoir at temperature $T_1$ or reservoir $T_2$.  The probability    this happens is $W^1_{\epsilon_j, \epsilon_i}dt$ and $W^2_{\epsilon_j, \epsilon_i}dt$ respectively. Note that the probability of a  simultaneous exchange of energy from both reservoirs goes as $dt^2$ and can be neglected. Hence the probability to make a jump $P(\epsilon_i \rightarrow \epsilon_j)dt$ is
\begin{eqnarray}
P(\epsilon_i \rightarrow \epsilon_j)dt =(W^1_{\epsilon_j, \epsilon_i}+W^2_{\epsilon_j, \epsilon_i})P(\epsilon_i)dt = W_{\epsilon_j, \epsilon_i}P(\epsilon_i)dt
\end{eqnarray}
where $P(\epsilon_i)$ is the probability the system is in energy $\epsilon_i$. This leads to  the equation  

\begin{eqnarray}
 W_{\epsilon_i, \epsilon_j} = W^{1}_{\epsilon_i, \epsilon_j} + W^{2}_{\epsilon_i, \epsilon_j}
 \label{proof}
\end{eqnarray}

which is of  the form of Eq.\ref{sum}. The $W$'s are supposed to obey a local detailed balance equation
 \begin{eqnarray}
	\frac{W^{1,2}_{\epsilon_i, \epsilon_j} }{W^{1,2}_{\epsilon_j, \epsilon_i} } &&= e^{-\beta^{1,2} (\epsilon_i-\epsilon_j)} \nonumber\\
	%\sum_{i=1,3} W_{\epsilon_i, \epsilon_j} &&=0, \quad each \; j
\end{eqnarray} 
%It is this local detailed balance that justifies ${=\sum_{k=1,n}\frac{\dot{Q}^k}{T^k} = -\dot{S}_{environment}}$ in Eq.\ref{2ndlaw}
%
%One could guess that   if $\epsilon_i> \epsilon_j$,   the rate  $W^1_{\epsilon_i, \epsilon_j}$ would represent the system taking energy from the bath. Since the bath is simply at temperature $T_1$ there should be no issue in the rate being dependent on temperature $T_1$. However, if $\epsilon_i< \epsilon_j$, then we have $W^1_{\epsilon_i, \epsilon_j}$  represents the system losing energy to the bath at temperature $T_1$.

However, now consider the limit $T_1 = T_2 = T$.  In this limit, you have the system is in contact with a thermal reservoir at temperature $T$ and the L.H.S in Eq.\ref{proof}  is then simply the transition rate for a system in contact with a reservoir at temperature $T$. But because $T_1 = T_2 = T$, we should also have that 	$W_{\epsilon_i, \epsilon_j} = W^1_{\epsilon_i, \epsilon_j} = W^2_{\epsilon_i, \epsilon_j}$. This leads to a contradiction in Eq.\ref{proof}. What has gone wrong here? One possible issue in the argument above is that detailed balance for a system at equilibrium with a bath only fixes the ratio between transition rates and not its magnitude. Hence,  $W_{\epsilon_i, \epsilon_j}$ when temperature $T_1 = T_2$ if is said to  equal to twice the transition rate for a system in contact with only one reservoir, we would lead to no contradiction with Eq.\ref{proof}. A thermal reservoir could be assumed to be a bath of particles at a particular temperature. The only way that two thermal reservoirs attached to the system would lead to twice the transition rate is if the thermal contact is a surface area contact between the bath and the system, so if two thermal reservoirs make contact with the system, it would imply twice the surface area contact, and   would hence imply twice the probability of particles of the bath interacting with the system leading to twice the transition rate. This is only possible if   that the system in question is  an extended system . In Section II, we study an extended system composed of multiple coupled degrees of freedom, where energy exchange with the environment occurs locally at different points of the system through contacts with thermal reservoirs at temperatures $T_1$ and $T_2$.
With such a construction we show that the evolution of the system as a whole is not Markovian in section II, implying possibilities for new research, such as understanding of fluctuation theorems etc for these systems.  In section III  we get away from considering a specific system and  consider a generic situation assuming Eq.\ref{proof} to be  true and assuming a steady state  exists, we show that one leads to extra constraints on the transition rates, implying a steady state generically is not possible. 

 {The results of this work fall into two logically distinct parts. In Section II, we show that coarse-graining a composite system interacting locally with multiple reservoirs generally leads to non-Markovian dynamics for the total energy, invalidating the usual Markovian interpretation of additive transition rates. In Section III, we go further and show that even if one nevertheless assumes a Markovian master equation with additive rates, the existence of a steady state leads to an explicit inconsistency.}

 %(We note that      despite observations of papers such as \cite{steadystate} a Markovian evolution with conserved probabilities need not imply a steady state, which we  explicitly show   in Appendix C.)  

%
%
%We  prove this fact in section II, by first considering a simple proof by setting all thermal reservoirs to be at the same temperature and illustrating an obvious inconsistency.In the same section we then show that even when the thermal reservoirs are not at the same temperature  Eq.\ref{sum} leads to extra constraints on the $W^k_{\epsilon_i, \epsilon_j}$ which is unduly restrictive.  In section III, we try to understand such systems connected to multiple reservoirs, by considering the simplest realization as  a system made up of two points connected to two different thermal reservoirs, with each point evolving in a Markovian fashion. Considering the two points as a single system, we show that the coarse grained evolution of the system generically cannot be Markovian. This result finds a natural extension to generic systems connected to two thermal reservoirs.   We   conclude by highlighting the difficulty in framing a second law of thermodynamics, if we  only use the information contained in $P(\epsilon)$. 

  \vspace{20pt}
  
 \section{  Two thermal reservoirs connected to an extended body }

In this section, we examine the Markovianity of the \emph{coarse-grained} dynamics of a system connected to multiple thermal reservoirs. Our analysis does \emph{not} rely on any assumption regarding the relative magnitudes of internal transition rates and reservoir-induced transition rates. %In particular, we do \emph{not} assume instantaneous equilibration within the system, nor do we assume any separation of time scales.

The only assumptions entering the discussion are:
\begin{enumerate} 
	\item energy exchange with each thermal reservoir occurs locally and satisfies local detailed balance at the corresponding temperature, and
	\item the full microscopic dynamics of the composite system is Markovian when resolved at the level of all subsystem energies.
\end{enumerate}

We emphasize that the question addressed here is not whether a specific physical limit admits an effective Markovian description, but whether the commonly used \emph{additive form of transition rates for the total energy} follows generically upon coarse-graining, independent of additional assumptions beyond those stated above.

\begin{figure}
 	\centering
 	\includegraphics[width=0.4\linewidth]{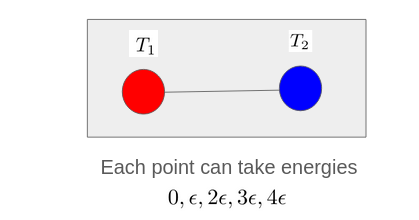}
 	\caption{The two point system made up of two points at different temperatures, which is studied in the 'Illustrative Example' section below.}
 	\label{example}
 \end{figure}

To isolate this issue, we consider the minimal setting in which two degrees of freedom of a single composite system are locally coupled to thermal reservoirs at different temperatures.
While the full dynamics of the joint system $(E_1,E_2)$ is Markovian by construction, we show below that the induced dynamics of the total energy $E = E_1 + E_2$ does \emph{not}, in general, admit a closed Markovian description. This obstruction arises purely from coarse-graining in the presence of incompatible local detailed-balance conditions, and is independent of any hierarchy of transition rates. Consider the example system shown in Fig.\ref{example}, where the energy at points $1$ and $2$ take only the following values   $ 0, \epsilon, 2\epsilon, 3 \epsilon,  4 \epsilon$.  
 
 %In calculations below the probabilities are assumed to be dependent on time.  The systems can exchange energy but are non-interacting   \footnote{It should be noted that this assumption is done in statistical mechanics to prove that two systems exchanging energy without interaction, reach a thermal equilibrium when both are at constant temperature. See for example Sec 4.1  in \cite{kardar}}. 
 
Let  system~1 have energy $E_1 = \epsilon_1$ and system~2   have energy $E_2 = \epsilon_2$. 
 If we label the  probability of this joint configuration as $P(E_1 = \epsilon_1, E_2 = \epsilon_2)$, then in the   limit $N \to \infty$, the number of ensembles with energies   $(E_1, E_2) = (\epsilon_1, \epsilon_2)$ is $N P(E_1 = \epsilon_1, E_2 = \epsilon_2)$.   Because the probability that system~1 makes a transition from energy $\epsilon_1$ to $\epsilon'_1$ in time $dt$, given by $w^1_{\epsilon'_1,\epsilon_1}\,dt$, then  by definition, the total number of   transitions among the ensembles during time $dt$ when subsystem~1 undergoes a transition $\epsilon_1 \rightarrow \epsilon'_1$  , while subsystem~2 remains at $E_2 = \epsilon_2$ is 
 
\begin{eqnarray}
 w^1_{\epsilon'_1, \epsilon_1}dt\, \times \, N\,    P(E_1 = \epsilon_1, E_2 = \epsilon_2).
\end{eqnarray}

 In the discussion below we represent, $p_1(\epsilon_1)p_2(\epsilon_2)$ as the joint probability $P(E_1=\epsilon_1,E_2=\epsilon_2)$ of the two subsystems, 
   purely for convenience; the discussion below does not rely on assuming any separability or factorization of the stationary distribution.
 
 To move forward let us first assume that the evolution as a whole is Markovian for all values of $T_1, T_2$.  All probabilities are assumed to be time dependent in the presentation below. The probability for the whole system to make a jump from total energy $5 \epsilon$ to $3\epsilon$ in time $dt$  is
 \begin{eqnarray}
 P(5\epsilon \rightarrow 3 \epsilon)dt &&= [w^1_{\epsilon, 3 \epsilon} + w^2_{0, 2 \epsilon}]dt p^1(3\epsilon)p^2(2\epsilon)  \nonumber\\
 &&  + [w^2_{\epsilon, 3 \epsilon} + w^1_{0, 2 \epsilon}] dt p^1(2\epsilon)p^2(3\epsilon)    \nonumber\\
 &&+ w^1_{2\epsilon, 4 \epsilon}dt p^1(4\epsilon)p^2(\epsilon)+ w^2_{2\epsilon, 4 \epsilon}dt p^1(\epsilon)p^2(4\epsilon)\nonumber\\
  &&+ w^1_{3\epsilon, 5 \epsilon}dt p^1(5\epsilon)p^2(0)+ w^2_{3\epsilon, 5 \epsilon}dt p^1(0)p^2(5\epsilon)\nonumber\\
 &&= WW^{1,2}_{3\epsilon, 5\epsilon}P(5\epsilon) dt
 \label{3e}
 \end{eqnarray}

 Also the probability to make the reverse jump from total energy $ 3 \epsilon$ to $5 \epsilon$ assuming a Markovian evolution is 
 \begin{eqnarray}
 P(3\epsilon \rightarrow 5 \epsilon) dt &&=    [w^2_{3\epsilon, \epsilon} +w^1_{4\epsilon, 2\epsilon}]dt p^1(2\epsilon)p^2( \epsilon)\nonumber\\
 &&+     [w^1_{3\epsilon,\epsilon} + w^2_{4\epsilon,2\epsilon}]dt  p^1(\epsilon)p^2(2\epsilon) \nonumber\\
 &&+    [ w^2_{2\epsilon,0}  + w^1_{5\epsilon,3\epsilon}] dt p^1(3\epsilon)p^2(0)\nonumber\\      &&+[w^1_{2\epsilon,0}    
  + w^2_{5\epsilon,3\epsilon}]dt p^1(0)p^2(3\epsilon)\nonumber\\
 &&=WW^{1,2}_{5\epsilon, 3\epsilon}P(3\epsilon) dt
 \label{5e}
 \end{eqnarray}
%In the above there is no requirement that the $W$'s have to be time independent. 

%  $ = WW^{1,2}_{3\epsilon, 5\epsilon}dt[ p^1(2\epsilon)p^2( 3\epsilon)+ p^1(3\epsilon)p^2(2\epsilon) +p^1(5\epsilon)p^2(0)+p^1(0)p^2(5\epsilon) +p^1(4\epsilon)p^2( \epsilon)+ p^1(\epsilon)p^2(4\epsilon)  ]$
 \subsubsection{ $\mathbf{ T_1 = T_2 = T }$}

Now consider the case where  both points are at the same temperature, i.e. $T_1 =T_2 = T$ . We note that, since, the system as a whole is at constant temperature $T$, the evolution has to be Markovian, so that an thermal equilibrium is reached. We have that  $w^{1 }_{\epsilon,\epsilon'}= w^{  2}_{\epsilon,\epsilon'}= w_{\epsilon,\epsilon'}$, $WW^{1,2}_{\epsilon,\epsilon'} = WW_{\epsilon,\epsilon'}$, both dependent on $T$ and  {
$P(E_1=\epsilon_1,E_2=\epsilon_2)=P(E_1=\epsilon_2,E_2=\epsilon_1)$  on grounds of symmetry, represented simply as $p(\epsilon_1)p(\epsilon_2)$ and/or $p(\epsilon_2)p(\epsilon_1)$ below for notational convenience.}

%   $p^1(\epsilon) = p^2(\epsilon) = p(\epsilon)$. We note that the $p^1(\epsilon), p^2(\epsilon)$
%are generically dependent on time and their equality is guaranteed on grounds of symmetry. 
 
  We now derive explicit constraints on $w_{\epsilon,\epsilon'}$ by substituting expression for $P(3\epsilon)$ and $P(5\epsilon)$ into Eq.\ref{3e} and Eq.\ref{5e}
 \begin{itemize}
  
 \item Substituting $P(3\epsilon) = 2[p(3\epsilon)p(0) + p(2\epsilon)p(\epsilon)]$ in Eq.\ref{5e} when $T_1 = T_2 = T$ gives
 \begin{eqnarray}
 	P(3\epsilon \rightarrow 5 \epsilon) dt &&=     
 	   2   [w_{3\epsilon,\epsilon} + w_{4\epsilon,2\epsilon}]dt  p(\epsilon)p(2\epsilon) \nonumber\\
  &&+2[w_{2\epsilon,0}    
 	+ w_{5\epsilon,3\epsilon}]dt p(0)p(3\epsilon)\nonumber\\
 	&&=2 WW_{5\epsilon, 3\epsilon}  [p(3\epsilon)p(0) + p(2\epsilon)p(\epsilon)]  dt
 	\label{7e}
 \end{eqnarray}
 
 The only way this relationship is true for all values of $\epsilon$ is if  $w_{\epsilon,\epsilon'}$ is    a function of $ \epsilon-\epsilon'$,  or
 \begin{eqnarray}
 	w_{\epsilon,\epsilon'}&&= f(\epsilon-\epsilon') \nonumber\\
 	&&	\; if \; \epsilon > \epsilon' 
 	\label{fs2}
 \end{eqnarray}
 $ \epsilon > \epsilon' $ is present above as the $ 	w_{\epsilon,\epsilon'}$ in Eq.\ref{7e} are of this form. 
 
\item Substituting   $P(5\epsilon) = 2[p(5\epsilon)p(0) + p(4\epsilon)p(\epsilon)+ p(3\epsilon)p(2\epsilon)]$ in Eq.\ref{3e}  when $T_1 = T_2 = T$ gives
 
 \begin{eqnarray}
 	P(5\epsilon \rightarrow 3 \epsilon)dt &&= 2[w_{\epsilon, 3 \epsilon} + w_{0, 2 \epsilon}]dt p(3\epsilon)p(2\epsilon)  \nonumber\\
 	&&+ [w_{2\epsilon, 4 \epsilon} + w_{2\epsilon, 4 \epsilon}]dt p(\epsilon)p(4\epsilon)\nonumber\\
 	&&+ [w_{3\epsilon, 5 \epsilon} + w_{3\epsilon, 5 \epsilon}]dt p(0)p(5\epsilon)\nonumber\\
 	&&= 2WW_{3\epsilon, 5\epsilon}[p(5\epsilon)p(0) + p(4\epsilon)p(\epsilon)+ p(3\epsilon)p(2\epsilon)] dt
 	\label{8e}
 \end{eqnarray}
 
 Unlike the case above we are now having $ 	w_{\epsilon,\epsilon'}$ with $ \epsilon' < \epsilon$. The only way this relationship is true for all values of $\epsilon$ is if $2[w_{\epsilon, 3 \epsilon} + w_{0, 2 \epsilon}] =  [w_{2\epsilon, 4 \epsilon} + w_{2\epsilon, 4 \epsilon}]= [w_{3\epsilon, 5 \epsilon} + w_{3\epsilon, 5 \epsilon}]$. If this is true for every value of $\epsilon$ we should have 
\end{itemize}

 \begin{eqnarray}
%w_{\epsilon,\epsilon'} &&= f(\epsilon-\epsilon')\nonumber\\
%&& \; if \; \epsilon>\epsilon' \nonumber\\
w_{\epsilon,\epsilon'} &&= f(\epsilon-\epsilon')\nonumber\\
&& \; if \; \epsilon'>\epsilon\; \nonumber\\
&& \; \epsilon'  < \frac{E + \epsilon}{2}   \nonumber\\
w_{\epsilon,\epsilon'} &&= 2f(\epsilon-\epsilon')\nonumber\\
&& \; if \; \epsilon'>\epsilon \;\nonumber\\
&& \; \epsilon'  > \frac{E + \epsilon}{2} .   \nonumber\\
\label{fs}
 \end{eqnarray}
 Here, $E$ is the total   energy of the system before making a transition and appears if a system makes a transition to a lower energy state.  
 
Combining the constraints from Eqs. \ref{fs2} and \ref{fs} leads to a compact expression for the coarse-grained transition rate  
 \begin{eqnarray}
WW_{E,E'} = 2f(E-E')
 \end{eqnarray}
 
 We hence see that the transition rate for the combined system is also a function of $E-E'$. We hence see the consistency of our approach. Each system $1$ and $2$ as well as the combined system at temperature $T$ as far as thermodynamics goes are simply exchanging energy (in their respective reference frame) with a thermal reservoir at temperature $T$, hence their functional dependence on energy levels between which they transition should be similar.

Since the entire system is exchanging energy with a a single thermal reservoir, detailed balance has to be obeyed. We get  
 \begin{eqnarray}
 \frac{f(E-E')}{f(E'-E)} = \frac{P_{eq}(E)}{P_{eq}(E')} =e^{-\beta (E-E')}.
 \label{detailed_balance}
 \end{eqnarray}
% We also find from Eq.\ref{fs} that $\frac{w_{\epsilon,\epsilon'}}{w_{\epsilon',\epsilon}}$ need not equal $  e^{-\beta (\epsilon-\epsilon')}$. 

\subsubsection{$\mathbf{T_1 \neq T_2}$}
Now, let us use what was derived above to analyze the case when $T_1 \neq T_2$. Using   Eq.\ref{5e} we get,
 \begin{eqnarray}
P(3\epsilon \rightarrow 5\epsilon)dt&&=  [f^1(2\epsilon) + f^2(2\epsilon)]dt P(3\epsilon),
 \end{eqnarray}
 however using Eq.\ref{3e} we have 
  \begin{eqnarray}
&& P(5\epsilon \rightarrow 3 \epsilon) dt = [f^1(-2\epsilon) + f^2(-2\epsilon)]dt p^1(3\epsilon)p^2(2\epsilon) \nonumber\\
 &&   +[f^1(-2\epsilon) + f^2(-2\epsilon)]dt p^1(2\epsilon)p^2(3\epsilon)    \nonumber\\
 &&+ 2 f^1(-2\epsilon)dt p^1(4\epsilon)p^2(\epsilon)+  2 f^2(-2\epsilon)dt p^1(\epsilon)p^2(4\epsilon)\nonumber\\
 &&+ 2 f^1(-2\epsilon)dt p^1(5\epsilon)p^2(0)+  2 f^2(-2\epsilon)dt p^1(0)p^2(5\epsilon)\nonumber\\
% &&= [f^1(-2\epsilon) + f^2(-2\epsilon)]dt p^1(3\epsilon)p^2(2\epsilon)    +[f^1(-2\epsilon) + f^2(-2\epsilon)]dt p^1(2\epsilon)p^2(3\epsilon)    \nonumber\\
% &&+  [f^1(-2\epsilon) + f^2(-2\epsilon)] p^1(4\epsilon)p^2(\epsilon)+  [f^1(-2\epsilon) + f^2(-2\epsilon)] p^1(\epsilon)p^2(4\epsilon)\nonumber\\
% &&+  [f^1(-2\epsilon) - f^2(-2\epsilon)] dt p^1(4\epsilon)p^2(\epsilon)+  [f^2(-2\epsilon) - f^1(-2\epsilon)]dt p^1(\epsilon)p^2(4\epsilon)\nonumber\\
 &&=[f^1(-2\epsilon) + f^2(-2\epsilon)] dt P(5\epsilon)\nonumber\\
 &&+  [f^1(-2\epsilon) - f^2(-2\epsilon)]dt[ p^1(4\epsilon)p^2(\epsilon) -p^1(\epsilon)p^2(4\epsilon)]\nonumber\\
 &&+  [f^1(-2\epsilon) - f^2(-2\epsilon)]dt[ p^1(5\epsilon)p^2(0) -p^1(0)p^2(5\epsilon)]\nonumber\\
 &&< 2 [f^1(-2\epsilon) + f^2(-2\epsilon)] dt P(5\epsilon)\nonumber\\
 \label{inequality1}
 \end{eqnarray}
 implying $ P(5\epsilon \rightarrow 3 \epsilon) dt$ cannot be written in a form $ WW^{1,2}(3\epsilon, 5\epsilon)dt P(5\epsilon)$, implying the system as a whole cannot evolve in a Markovian fashion.  We hence see that rightly assuming Markovian evolution when $T_1 = T_2$, directly leads to non Markovian dynamics for $T_1 \neq T_2$.  {This shows that the additive-rate form implicitly assumes a Markovian coarse-grained description which is not generically valid once the system interacts locally with multiple reservoirs.}

 We   generalize   this observation and claim  that if the minimum value of energy at a particular site is $\epsilon_{min}$, then if $E > E'$ and $E-E'> \epsilon_{min}$ then 
\begin{eqnarray}
 P(E \rightarrow E')dt&&< 2 [f^1(E'-E) + f^2(E'-E)] dt P(E)   \nonumber\\
 P(E'\rightarrow E) dt&&= [f^1(E-E') + f^2(E-E')] dt P(E')   \nonumber\\
 \label{inequality_1}
\end{eqnarray}

\subsection{Extended body}
We can model an extended body by considering points $1$ and $2$ in contact with other points we label as $a, b, c..$ etc that are themselves not in contact with a thermal reservoir. Then the probability of the system having energy $E$ is 
\begin{eqnarray}
	P(E)=\sum_{\epsilon_1, \epsilon_2, \epsilon_a...} p^1(\epsilon_1) p^2(\epsilon_2)p^a(\epsilon_a)... \delta_{\epsilon_1+ \epsilon_2+ \epsilon_a + ... , E}.\nonumber\\
\end{eqnarray}
Since energy exchanges with the environment still happen at points $1$ and $2$,    the evaluations of $P(E\rightarrow E')$ in this case also follow everything said up until  Eq.\ref{inequality_1}, implying a non-Markovian  evolution.

\subsection{Issues with deriving Second Law.}
 {We emphasize that the observations below do not constitute a violation of the second law, but rather reflect the inapplicability of standard Markovian entropy production formulas when the coarse-grained dynamics is non-Markovian.} We now begin to understand the difficulty in deriving a second law of thermodynamics using $S_{system}=-k_B \sum_E P(E)\ln P(E)$ for this problem. For any $E,E'$ such that $E>E'$ we could atleast write 
\begin{eqnarray}
P(E' \rightarrow E)dt&&= [w^1_{E,E'} + w^2_{E,E'}] dt P(E'), \quad  \nonumber\\
\end{eqnarray}
while for $E<E'$ we get an inequality  Eq.\ref{inequality_1}. However, the only   inequality that would have guaranteed the  second law  is  instead
\begin{eqnarray}
P(E' \rightarrow E)dt&&\geq [w^1_{E,E'} + w^2_{E,E'}] dt P(E'), \quad if \; E < E' \nonumber\\
\label{inequality}
\end{eqnarray}
along with $\frac{w_{E,E'}}{w_{E',E}}= e^{-\beta (E-E')}$ as is shown in literature \cite{stochastic1}.  However  Eq.\ref{fs} and Eq.\ref{detailed_balance} tells us that  $\frac{w_{E,E'}}{w_{E',E}}$ need not equal $  e^{-\beta (E-E')}$.

%This is because in such a case we have
%\begin{eqnarray}
%\dot{S }_{system} &&= - k_B \sum_E d_t P(E)\ln P(E) = k_B  \sum_{E,E'}[P(E\rightarrow E')- P(E'\rightarrow E)]\ln P(E) \nonumber\\
%&&= \frac{k_B}{2 }\sum_{E,E'} [P(E\rightarrow E') - P(E'\rightarrow E)]  \ln\frac{P(E)}{P(E')}\nonumber\\
%&&> \frac{k_B}{2}\sum_{E,E',\nu\in \{1,2\}} [W^\nu_{E',E}  P(E)  - W^\nu_{E,E'} P(E') ]\ln\frac{P(E)}{P(E')}\nonumber\\
%&&=\underbrace{\frac{k_B}{2}\sum_{E,E',\nu\in \{1,2\}}[W^\nu_{E',E}  P(E)  - W^\nu_{E,E'} P(E') ]\ln \frac{W^\nu_{E',E}  P(E)}{ W^\nu_{E,E'} P(E') }}_{>0}  \nonumber\\
%&&+\underbrace{\frac{k_B}{2} \sum_{E,E',\nu\in \{1,2\}}[W^\nu_{E',E}  P(E)  - W^\nu_{E,E'} P(E') ] \ln \frac{W^\nu_{E,E'}   }{ W^\nu_{E',E} } }_{=\sum_{\nu=1,2}\frac{dQ^\nu}{T^\nu} = -\dot{S}_{environment}}  \nonumber\\
%\label{2ndlaw}
%\end{eqnarray}
    To understand why $S_{system}=-k_B \sum_E P(E)\ln P(E)$ is not the 'right' entropy for a second law calculation,   note that at first 

\begin{eqnarray}
 {S}_{system} &&= -k_B \sum_E   P(E)\ln P(E)\nonumber\\
  &&=-k_B \sum_E   \sum_\epsilon p^1(E-\epsilon)p^2(\epsilon)\ln  \sum_{\epsilon'} p^1(E-\epsilon')p^2(\epsilon') \nonumber\\
&&< -k_B     \sum_{E,\epsilon} p^1(E-\epsilon)p^2(\epsilon) \ln    p^1(E-\epsilon  )p^2(\epsilon  )\nonumber\\
&& = {S}_{point \; 1 } +{S}_{point \; 2 } \nonumber\\
\end{eqnarray}
 where ${S}_{point \; 1 },{S}_{point \; 2 }$ are entropies of points $1$ and $2$ respectively. 
Looking from the lens of information theory, this implies that  , $S_{system}$ doesn't contain  the total information content of points $1$ and $2$.

 Now, consider a system of $N$ particles in contact with a thermal bath at temperature $T$. Let us say the number of particles having energy   $\epsilon_i$ is $n_i$. Then the total number of ways of realizing this arrangement is given by 
$\frac{N!}{\Pi_i n_i!}$ which in limit of large $N$  is $ e^{-N\sum_i \frac{n_i}{N}\ln \frac{n_i}{N}} = e^{-N\sum_i p_i \ln p_i}$, where $p_i = \frac{n_i}{N}$ is the probability of  finding particles with energy $\epsilon_i$. Since $-k_B\sum_i p_i \ln p_i$ is the entropy per particle we have that $S_{particles} = - k_B N\sum_i p_i \ln p_i$ is the total entropy of the particles.  If this changes in time evolution by $\triangle S_{particles}$  and since  $\triangle S_{environment} =  \frac{\triangle Q}{T}$ is the change in the entropy of the environment, where $\triangle Q $ is heat gained by the enivonment, then the second law stating that $\triangle S_{particles} + \triangle S_{environment} > 0$ is the statement that the system plus the environment evolves in to a macrostate that has maximum possible realizations. Hence, since all particles are independent of each other the relation $\frac{ \triangle S_{particles} + \triangle S_{environment}  }{N}>0 $, says that total change in entropy per particle and the entropy change in environment because of heat released by this particle on an average should be greater than zero. For Markovian systems this is can be easily derived \cite{stochastic1}.  Second law  hence argues that the environment plus the system moves towards the macrostate which has maximum possible realizations. It is because the  number of ways of realizing a macrostate   is of relevance to the second law, that  the relevant entropy of composite system of points $1$ and $2 $ should be ${S}_{point \; 1 } +{S}_{point \; 2 }  $ and not $S_{system}$. $  {S}_{point \; 1 } + {S}_{point \; 2 } +  S_{environment} $   will be increasing with time as each point is in contact with a thermal bath and the  entropy change of each point plus the energy it releases into the environment increases as per arguments in this paragraph and standard derivations in literature  \cite{stochastic1}.  Hence, $S_{system}=-k_B \sum_E P(E)\ln P(E)$ is not the 'right' entropy for a second law calculation.

\section{      $  W_{\epsilon_i, \epsilon_j} = W^1_{\epsilon_i, \epsilon_j} + W^2_{\epsilon_i, \epsilon_j}$ and existence of  steady state leads to a contradiction. } 

Consider a system that can have three possible energies labelled as $\epsilon_i$ with $i \in [1,2,3]$. We also assume that system   follows a Markovian evolution given by Eq.\ref{master_equation} 	Let us assume that the system is in contact with two thermal reservoirs with temperatures $T_1$ and $T_2$ and as quoted in literature let us assume that   the transition rate obeys

\begin{eqnarray}
	W_{\epsilon_i, \epsilon_j} = W^1_{\epsilon_i, \epsilon_j} + W^2_{\epsilon_i, \epsilon_j}.
	\label{sum_2}
\end{eqnarray}  

%Next, assume the local detailed balance below, along with  the   fact that the steady state exists, i.e. Eq.\ref{steady-state} is true for any $T_1$, $T_2$. Then doing some algebra after Eq.\ref{steady-state} we arrive at  equation Eq.\ref{Eq17}, where the   LHS depends on temperature $T_1$ and the RHS on temperature $T_2$, implying these have to be constant. This implies an extra constraint on the $W$'s other than a local detailed balance, which is inconsistent.  
%
%After summarizing how the proof goes in paragraph above, let us go through the actual proof.  

The local detailed balance relationships are 
\begin{eqnarray}
	\frac{W^{1,2}_{\epsilon_i, \epsilon_j} }{W^{1,2}_{\epsilon_j, \epsilon_i} } &&= e^{-\beta^{1,2} (\epsilon_i-\epsilon_j)} \nonumber\\
	%\sum_{i=1,3} W_{\epsilon_i, \epsilon_j} &&=0, \quad each \; j
	\label{constraints}
\end{eqnarray} 
we can hence write in the steady state
\begin{eqnarray}
	&&\sum_{ \epsilon_j, \epsilon_j \neq \epsilon_i  } [ W^1_{\epsilon_j, \epsilon_i} e^{-\beta^{1 } (\epsilon_i-\epsilon_j)}  + W^2_{\epsilon_j, \epsilon_i} e^{-\beta^{ 2} (\epsilon_i-\epsilon_j)} ] P_{steady}(\epsilon_j) \nonumber\\
	&&=   \sum_{ \epsilon_j, \epsilon_j  \neq \epsilon_i} [ W^1_{\epsilon_j, \epsilon_i} + W^2_{\epsilon_j, \epsilon_i}] P_{steady}(\epsilon_i). \nonumber\\
	\label{steady-state}
\end{eqnarray}

This equation is written explicitly by substituting the detailed-balance expressions of Eq. \ref{constraints} into the master equation. Note that Eq.\ref{steady-state} is  the complete steady-state master equation, not a simplification of Eq. \ref{constraints}. Even though there are three equations above corresponding to $i= [1,3]$, they are not independent, with one of them being a linear combination of the other two. 
%	Because of the symmetry   we see that $P(\epsilon_i)$ should be a symmetric function of  $T_1$ and $T_2$.  
 Now if we consider a column vector  $W^{1,2}$  
\begin{align*} W^{1,2}= 
	\begin{bmatrix}
		W^{1,2}_{\epsilon_1, \epsilon_2} \\ 
		W^{1,2}_{\epsilon_2, \epsilon_3}  \\
	%	W^{1,2}_{\epsilon_3, \epsilon_1}   \\ 
	\end{bmatrix},  
\end{align*}

%Here, we simply collect all independent elements of the 
%$3\times3$ transition-rate matrices $W^1$ and $W^2$ into 
%column vectors $W^1, W^2$ using the standard vectorization 
%convention, so that subsequent equations can be written in 
%compact matrix form. No information is lost --- this is purely 
%a notational simplification.

then the above equations can be written in a matrix form as 

\begin{eqnarray}
F(\beta_1, \beta_2) +	N(\beta_1, \beta_2)W^1 + F(\beta_2, \beta_1)+N(\beta_2, \beta_1)W^2 = 0.
	\label{eq8}
\end{eqnarray}	
Here $N(\beta_1, \beta_2) $ is a $2 \times 2$ matrix, $F(\beta_1, \beta_2)$ is a column vector and $N(\beta_2, \beta_1)/ F(\beta_2, \beta_1)$ are obtained from $N(\beta_1, \beta_2)/F(\beta_1, \beta_2) $ by interchanging $\beta_1$ and $\beta_2$. We note that the  dependence of matrix $N(\beta_1,\beta_2)/F(\beta_1, \beta_2)$ on both $\beta_1, \beta_2$ is because the $P_{steady}$'s   are dependent on both $\beta_1, \beta_2$. The explicit forms of  $N(\beta_1,\beta_2)/F(\beta_1, \beta_2)$ are  given in appendix A.

%Eq.\ref{steady-state} represents the full steady-state master equation for the three-level system,
%written explicitly after substituting the detailed-balance relations of Eq. \ref{constraints}.
%The matrix $M(\beta_1,\beta_2)$ introduced in Eq. \ref{eq8} collects the coefficients of the
%transition rates $W^1$ and $W^2$ from this steady-state condition; its explicit form is  given in Appendix A.

We next note that the   $N(\beta_1, \beta_2)/F(\beta_1, \beta_2)$ are continous functions of $\beta_1, \beta_2$ and in the limit $\beta_1 = \beta_2 = \beta$
\begin{eqnarray}
	N(\beta  , \beta  )   &&=   0\nonumber\\
	F(\beta,\beta)&&=0.
	\label{equation}
\end{eqnarray}	
This is because when  $\beta_1 = \beta_2 = \beta$, the steady state is the thermal equilibrium state and $P(\epsilon_i) \propto e^{-\beta \epsilon_i}$.

For $\beta_1 \neq \beta_2$ we get from Eq.\ref{eq8}
\begin{eqnarray}
	W^1  &=& A(\beta_1, \beta_2) W^2 + B(\beta_1, \beta_2)\nonumber\\
	\label{eq7}
\end{eqnarray}
where 
\begin{eqnarray}
	A(\beta_1, \beta_2)&&= -	N(\beta_1, \beta_2)^{-1}N(\beta_2,\beta_1)\nonumber\\
	B(\beta_1, \beta_2)&&=-	N(\beta_1, \beta_2)^{-1}[F(\beta_1, \beta_2) +F(\beta_2, \beta_1) ]
	\label{limit}
\end{eqnarray}

\begin{comment}	
	Note   despite  Eq.\ref{equation}, we see from Eq.\ref{eq7} that $A(\beta_1, \beta_1) $ is the unit matrix. We could say that this is how the limit $\beta_1 \rightarrow \beta_2$ is taken in Eq.\ref{limit}.
\end{comment}
The above equation assumes existence of the inverse of matrix $ N(\beta_1, \beta_2)$, which   exists as shown in Appendix A.

% This requires the matrix to be   of full rank. To show this, we will assume the matrix  $ M(\beta_1, \beta_2)$ is not of full rank and show it leads to a contradiction.  Since,
%$\sum_i P_{steady}(\epsilon_i)=1$,   one of the 
%probabilities can be expressed in terms of other two. Assume that we express $ P_{steady}(\epsilon_1)$   in terms of  $ P_{steady}(\epsilon_2)$ and $ P_{steady}(\epsilon_3)$. If the matrix $M(\beta_1, \beta_2)$ is not full rank, it implies that the  determinant of $M(\beta_1, \beta_2)$ equals zero. However since $M(\beta_1, \beta_2)$ depends on both $ P_{steady}(\epsilon_2)$ and $ P_{steady}(\epsilon_3)$ by construction, it would imply  a constraint between $ P_{steady}(\epsilon_2)$ and $ P_{steady}(\epsilon_3)$. This constraint would be  independent of the $W$'s, as the matrix $ M(\beta_1, \beta_2)$ is itself independent of the $W$'s. This      would then imply that $ P_{steady}(\epsilon_2)$ and $ P_{steady}(\epsilon_3)$ are related to each other through a relationship independent of the $W$'s ( and hence the system under consideration)   which is not possible for generic values of $W$'s. We have hence proven that the matrix $ M(\beta_1, \beta_2)$  is of full rank and hence Eq.\ref{limit} is consistent.   

Now Eq.\ref{eq7} is an algebraic relationship, true for any values of the $\beta$'s. Hence if a system was in contact with two thermal reservoirs at temperatures $T_2, T_3$ we would  have 
\begin{eqnarray}
	W^2  &=& A(\beta_2, \beta_3) W^3+ B(\beta_2, \beta_3).\nonumber\\
	\label{eq27}
\end{eqnarray}
%for $\beta_1 \neq \beta_2$ 	  and $\beta_2 \neq \beta_3$
Substituting in Eq.\ref{eq7} gives
\begin{align}
	W^1
%	&= A(\beta_1,\beta_2)\big(A_{23} W^3 + B_{23}\big) + B_{12} \nonumber\\
	&= A(\beta_1,\beta_2)A(\beta_2, \beta_3)\, W^3 + A(\beta_1,\beta_2) B(\beta_2,\beta_3) + B(\beta_1,\beta_2).
	\label{eq:composition}
\end{align}
Since we also have
\begin{eqnarray}
W^1  &=& A(\beta_1, \beta_3) W^3 + B(\beta_1, \beta_3)\nonumber\\
\label{eq36}
\end{eqnarray}

we can show that assuming additive transition rates leads to a contradiction.
From Eqs.~(31) and (32), which must hold simultaneously, subtracting the latter
from the former yields a linear equation of the form
\begin{equation}
	M(\beta_1,\beta_2)\, W^{3}
	=
	C(\beta_1,\beta_2,\beta_3),
\end{equation}
where the matrix \(M\) depends only on \(\beta_1\) and \(\beta_2\), while the
right-hand side depends on \(\beta_1,\beta_2,\beta_3\).
Since the entries of \(M(\beta_1,\beta_2)\) are continuous functions of
\(\beta_1\) and \(\beta_2\) and there is no structural identity enforcing
\(M(\beta_1,\beta_2)\equiv 0\), it follows that for generic choices of
\((\beta_1,\beta_2)\) the matrix \(M\) is invertible.
Inverting \(M\) then yields an explicit expression for \(W^{3}\) that depends
on \(\beta_1\) and \(\beta_2\) in addition to \(\beta_3\).
This is a contradiction, since \(W^{3}\) is by definition the rate vector
associated with bath~3 and may depend on \(\beta_3\) but cannot depend on the
temperatures of other baths that are not present.
We therefore conclude that the assumption of additive transition rates,
together with local detailed balance and the existence of a steady state,
is inconsistent for a system with three energy levels.
The same reasoning extends straightforwardly to systems with more than three
levels.

\section*{IV. Discussion} 
The Markovian master equation  in Eq.\ref{master_equation} has a unique steady state solution $P_{steady}(\epsilon_i)$  such that

 \begin{eqnarray}
  \sum_{\epsilon_j} W_{\epsilon_i,\epsilon_j}P_{steady}(\epsilon_j) - \sum_{\epsilon_j} W_{\epsilon_j,\epsilon_i}P_{steady}(\epsilon_i) = 0\nonumber\\
\end{eqnarray}
 
 The uniqueness of $P_{steady}(\epsilon_i)$   is guaranteed by the linearity of the above set of  equations. The entropy production rate is written as \cite{seifert2}
\begin{eqnarray}
&&\frac{k_B}{2}\sum_{\epsilon_i, \epsilon_j} [W_{\epsilon_i, \epsilon_j}P_{steady}(\epsilon_j) - W_{\epsilon_j, \epsilon_i} P_{steady}(\epsilon_i)]\nonumber\\
 &&\times \ln \frac{ W_{\epsilon_i, \epsilon_j}P_{steady}(\epsilon_j)}{  W_{\epsilon_j, \epsilon_i} P_{steady}(\epsilon_i)}.\nonumber\\  
\label{entropy_production_rate}
\end{eqnarray}

We know that in case the system is connected to a single thermal reservoir, the steady state corresponds to the equilibrium state that obeys detailed balance and hence the entropy production at equilibrium is zero.  For a system connected to multiple reservoirs, be it thermal or particle reservoirs etc,    if a Markovian evolution is assumed  the above entropy production is greater than zero at the steady state and this is used to derive thermodynamic uncertainity relations \cite{seifert1}, \cite{seifert2}, \cite{currents_nature}
What we have shown in our work is that atleast for systems connected to multiple thermal reservoirs we cannot use Eq.\ref{sum} to evaluate the rate of entropy production. We have also shown that the evolution of such systems may not be Markovian hence Eq.\ref{entropy_production_rate} is not the rate of entropy production in the steady state.   Because of   non-Markovian evolution one cannot extend known ideas such as fluctuation theorems to such systems as a whole, despite the fact that parts of the system can individually satisfy fluctuation theorems. Since $S_{system}$ is not relevant to the second law, the question naturally arises if any other entropy definition that is dependent on probability distribution of the system as a whole could be used to construct a second law.  We also note that the environment for a particular point also includes the other point and hence the $S_{environment}$ as discussed in the above calculation does not consider just the energy exchanged by the composite system of points $1$ and $2$ with the environment, which is the universe minus the points $1$ and $2$, but also the energy exchanged by points $1$ and $2$ with each other. It is however, the universe minus the points $1$ and $2$ which would be of relevance if we were to frame a second law by constructing a entropy definition that only utilized $P(E)$. How  to accomplish this is an open question   and would require further research. 

%\section*{V. Acknowledgements}
%We would like to thank Prof Ranjan Mukhopadhyay for discussions on statistical thermodynamics. 
% 

 \appendix

 \section{Explicit form of $M(\beta_1, \beta_2)$}

 The transition rates were satisfying detailed balance with respect to inverse
 temperatures \(\beta_1,\beta_2\),
 \begin{equation}
 	\frac{W^{1,2}_{\varepsilon_i,\varepsilon_j}}{W^{1,2}_{\varepsilon_j,\varepsilon_i}}
 	=
 	e^{-\beta_{1,2}(\varepsilon_i-\varepsilon_j)} .
 \end{equation}
 
 We had defined
 \begin{equation}
 	W^{1,2}
 	=
 	\begin{pmatrix}
 		W^{1,2}_{\varepsilon_1,\varepsilon_2} \\
 		W^{1,2}_{\varepsilon_2,\varepsilon_3} \\
 	%	W^{1,2}_{\varepsilon_3,\varepsilon_1}
 	\end{pmatrix}.
 \end{equation}
 
We had the steady state master equation

\begin{eqnarray}
	&&\sum_{ \epsilon_j, \epsilon_j \neq \epsilon_i  } [ W^1_{\epsilon_j, \epsilon_i} e^{-\beta^{1 } (\epsilon_i-\epsilon_j)}  + W^2_{\epsilon_j, \epsilon_i} e^{-\beta^{ 2} (\epsilon_i-\epsilon_j)} ] P_{steady}(\epsilon_j) \nonumber\\
	&&=   \sum_{ \epsilon_j, \epsilon_j  \neq \epsilon_i} [ W^1_{\epsilon_j, \epsilon_i} + W^2_{\epsilon_j, \epsilon_i}] P_{steady}(\epsilon_i). \nonumber\\
\end{eqnarray}
 writing  $P_{steady}(\epsilon_i) = P_i$  we have

\ 
\begin{equation}
	\sum_{j\neq i}
	\Big[
	W^1_{\varepsilon_j,\varepsilon_i}
	e^{-\beta_1(\varepsilon_i-\varepsilon_j)}
	+
	W^2_{\varepsilon_j,\varepsilon_i}
	e^{-\beta_2(\varepsilon_i-\varepsilon_j)}
	\Big] P_j
	=
	\sum_{j\neq i}
	\Big[
	W^1_{\varepsilon_j,\varepsilon_i}
	+
	W^2_{\varepsilon_j,\varepsilon_i}
	\Big] P_i .
\end{equation}
%or
%
%\begin{equation}
%	\sum_{j\neq i}
%	\Big[
%	W^1_{\varepsilon_i,\varepsilon_j}
%	+
%	W^2_{\varepsilon_i,\varepsilon_j}
%	\Big] P_j
%	=
%	\sum_{j\neq i}
%	\Big[
%	W^1_{\varepsilon_j,\varepsilon_i}
%	+
%	W^2_{\varepsilon_j,\varepsilon_i}
%	\Big] P_i .
%\end{equation}
%\paragraph{Equation for \(i=1\).}

For \(i=1\), the sum runs over \(j=2,3\).
Choosing the independent rates
\(W_{1,2}\) and \(W_{3,1}\),
and eliminating reverse rates using detailed balance, we obtain
\begin{equation}
	\begin{aligned}
		0
		=&\;
		\Big( P_2 - P_1 e^{-\beta_1(\varepsilon_2-\varepsilon_1)} \Big)\, W^{1}_{12}
		+
		\Big( P_3 e^{-\beta_1(\varepsilon_1-\varepsilon_3)} - P_1 \Big)\, W^{1}_{31}+
		(1 \leftrightarrow 2 \text{ bath})
		\\
	\end{aligned}
\end{equation}

%\paragraph{Equation for \(i=2\).}

For \(i=2\), the sum runs over \(j=1,3\).
The independent rates are \(W_{1,2}\) and \(W_{2,3}\), yielding
\begin{equation}
	\begin{aligned}
		0
		=&\;
		\Big( P_1 e^{-\beta_1(\varepsilon_2-\varepsilon_1)} - P_2 \Big)\, W^{1}_{12}
		+
		\Big( P_3 - P_2 e^{-\beta_1(\varepsilon_3-\varepsilon_2)} \Big)\, W^{1}_{23}+
		(1 \leftrightarrow 2 \text{ bath})
		\\
	\end{aligned}
\end{equation}

%\paragraph{Equation for \(i=3\).}

For \(i=3\), the sum runs over \(j=1,2\).
\begin{equation}
	\begin{aligned}
		0
		=&\;
		\Big( P_1 - P_3 e^{-\beta_1(\varepsilon_1-\varepsilon_3)} \Big)\, W^{1}_{31}
		+
		\Big( P_2 e^{-\beta_1(\varepsilon_3-\varepsilon_2)} - P_3 \Big)\, W^{1}_{23}+
		(1 \leftrightarrow 2 \text{ bath})
		\\
	\end{aligned}
\end{equation}

One can see that the last equation can be got by summing the first two equations. Hence not all three equations are independent. 
%\subsection*{Matrix formulation}

 Once we consider the   steady--state condition in compact matrix form as in manuscript
\begin{eqnarray}
	F(\beta_1, \beta_2) +	N(\beta_1, \beta_2)W^1 + F(\beta_2, \beta_1)+N(\beta_2, \beta_1)W^2 = 0.
\end{eqnarray}	
the matrix \(N(\beta_1,\beta_2)\) is given explicitly by

\begin{equation}
	N(\beta_1,\beta_2)
	=
	\begin{pmatrix}
		P_2 - P_1 e^{-\beta_1(\varepsilon_2-\varepsilon_1)}
		&
		0
		\\[8pt]
		P_1 e^{-\beta_1(\varepsilon_2-\varepsilon_1)} - P_2
		&
		P_3 - P_2 e^{-\beta_1(\varepsilon_3-\varepsilon_2)}
		\\[8pt]
	\end{pmatrix}.
\end{equation}

\begin{equation}
	F(\beta_1,\beta_2)
	=W^1_{3,1}
	\begin{pmatrix}
		P_3 e^{-\beta_1(\varepsilon_1-\varepsilon_3)} - P_1
		\\[8pt]
		0
		\\[8pt]
	\end{pmatrix}.
\end{equation}

Since the explicit \(2\times2\) form of \(N(\beta_1,\beta_2)\) is given above,
it follows immediately that \(N(\beta_1,\beta_2)\) is invertible for generic
values of the parameters.

% \section{ }
% 	To see this note that the  rate of change of the system entropy is 
% 
% \begin{eqnarray}
% 	&&\dot{S}_{system} =\frac{d}{dt}[-k_B \sum_{\epsilon_i} P(\epsilon_i)\ln P(\epsilon_i)] =  -k_B \sum_{\epsilon_i} \frac{dP(\epsilon_i)}{dt} \ln P(\epsilon_i) \nonumber\\
% 	&&= -k_B \sum_{\epsilon_i, \epsilon_j}[ W_{\epsilon_i, \epsilon_j}P(\epsilon_j)-W_{\epsilon_j, \epsilon_i}P(\epsilon_i)]\ln P(\epsilon_i) \nonumber\\
% 	&&=\frac{k_B}{2} \sum_{\epsilon_i, \epsilon_j} [W_{\epsilon_i, \epsilon_j}P(\epsilon_j) -W_{\epsilon_j, \epsilon_i}P(\epsilon_i) ] \ln \frac{P(\epsilon_j)}{ P(\epsilon_i)}  \nonumber\\
% 	&&=\underbrace{\frac{k_B}{2} \sum_{k=1,n} \sum_{\epsilon_i, \epsilon_j} [W^k_{\epsilon_i, \epsilon_j}P(\epsilon_j) -W^k_{\epsilon_j, \epsilon_i}P(\epsilon_i) ] \ln \frac{W^k_{\epsilon_i, \epsilon_j}P(\epsilon_j)}{W^k_{\epsilon_j, \epsilon_i} P(\epsilon_i)}}_{\geq 0}  \nonumber\\
% 	&&+\underbrace{\frac{k_B}{2} \sum_{k=1,n} \sum_{\epsilon_i, \epsilon_j} [W^k_{\epsilon_i, \epsilon_j}P(\epsilon_j) -W^k_{\epsilon_j, \epsilon_i}P(\epsilon_i) ] \ln \frac{W^k_{\epsilon_j, \epsilon_i}  }{W^k_{\epsilon_i, \epsilon_j} } }_{=\sum_{k=1,n}\frac{\dot{Q}^k}{T^k} = -\dot{S}_{environment}}  \nonumber\\
% 	\label{2ndlaw}
% \end{eqnarray}
% Hence, we see that using Eq.\ref{sum}  we get the second law of thermodynamics $ \dot{S}_{system}+ 	\dot{S}_{environment}>0$. The above argument is taken from \cite{stochastic1}.
% 

\section*{Acknowledgements}
We would like to thank Karsten Kruse for  comments on the manuscript, Ranjan Mukhopadhyay for discussions on statistical thermodynamics and for comments on the manuscript,    M. Bhaskaran and Abitosh Upadhyay for going over a version of the proof in section III.

\section*{Data Availability}
There is no data associated with this research. 

\section*{Conflict of Interest}
There are no conflict of interests in this research as Vaibhav Wasnik is the sole author of this paper.


\begin{thebibliography}{9}
	\bibitem{stochastic1}
	 C. Van den Broeck and M. Esposito, Ensemble and
	trajectory thermodynamics: A brief introduction, Physica
	(Amsterdam) 418A, 6 (2015)
	\href{https://ui.adsabs.harvard.edu/link_gateway/2015PhyA..418....6V/doi:10.1016/j.physa.2014.04.035}{DOI: $
		10.1016/j.physa.2014.04.035
		$ }
	\bibitem{stochastic2}
	 C. Jarzynski, Equilibrium free-energy differences from
	nonequilibrium measurements: A master-equation ap-
	proach, Phys. Rev. E 56, 5018 (1997).
	\href{https://doi.org/10.1103/PhysRevE.56.5018}{DOI :$https://doi.org/10.1103/PhysRevE.56.5018$}
		\bibitem{stochastic3}
	G. E. Crooks, Entropy production fluctuation theorem
	and the nonequilibrium work relation for free energy
	differences, Phys. Rev. E 60, 2721 (1999).
	\href{DOI:https://doi.org/10.1103/PhysRevE.60.2721}{DOI: $https://doi.org/10.1103/PhysRevE.60.2721$}
		\bibitem{stochastic4}
	C. Jarzynski, Equalities and inequalities: Irreversibility and
	the second law of thermodynamics at the nanoscale, Annu.
	Rev. Condens. Matter Phys. 2, 329 (2011).
	\href{ https://doi.org/10.1146/annurev-conmatphys-062910-140506 }{DOI: $ https://doi.org/10.1146/annurev-conmatphys-062910-140506 $}
	\bibitem{stochastic5}
	  G. E. Crooks, Nonequilibrium measurements of free energy
	differences for microscopically reversible Markovian sys-
	tems, J. Stat. Phys. 90, 1481 (1998).
  \href{https://doi.org/10.1023/A:1023208217925 }{DOI: $ https://doi.org/10.1023/A:1023208217925$}
 \bibitem{stochastic6}
 U. Seifert, Stochastic thermodynamics, fluctuation theorems
 and molecular machines, Rep. Prog. Phys. 75, 126001
 (2012).
 \href{doi:10.1088/0034-4885/75/12/126001}{doi:10.1088/0034-4885/75/12/126001}
	\bibitem{stochastic7}
	 C. Maes, On the origin and the use of fluctuation relations
	for the entropy, Semin. Poincar e 2, 29 (2003).
	\bibitem{stochastic8}
	  K. Sekimoto, Stochastic Energetics (Springer, Berlin,
	Germany, 2010).
 \href{DOI 10.1007/978-3-642-05411-2.}{DOI 10.1007/978-3-642-05411-2.}
 
 
	\bibitem{demon1}
	Strasberg, P., Schaller, G., Brandes, T.,  $\&$  Esposito, M. (2013). Thermodynamics of a physical model implementing a Maxwell demon. Physical review letters, 110(4), 040601.
	\href{DOI:https://doi.org/10.1103/PhysRevLett.110.040601}{DOI:https://doi.org/10.1103/PhysRevLett.110.040601}
	\bibitem{demon2}
	Esposito, M., $\&$ Schaller, G. (2012). Stochastic thermodynamics for “Maxwell demon” feedbacks. EPL (Europhysics Letters), 99(3), 30003.
	\href{10.1209/0295-5075/99/30003}{DOI: 10.1209/0295-5075/99/30003}
	\bibitem{sum_1}
	Barato, A. C., $\&$ Seifert, U. (2014). Stochastic thermodynamics with information reservoirs. Physical Review E, 90(4), 042150.
	\href{DOI:https://doi.org/10.1103/PhysRevE.90.042150}{DOI: $https://doi.org/10.1103/PhysRevE.90.042150$ }
\bibitem{multiple1}
Esposito, M. (2012). Stochastic thermodynamics under coarse graining. Physical Review E, 85(4), 041125.
\href{DOI:https://doi.org/10.1103/PhysRevE.85.041125}{DOI : $  https://doi.org/10.1103/PhysRevE.85.041125$}
\bibitem{multiple2}
Proesmans, K., $\&$ Fiore, C. E. (2019). General linear thermodynamics for periodically driven systems with multiple reservoirs. Physical Review E, 100(2), 022141.
\href{DOI:https://doi.org/10.1103/PhysRevE.100.022141}{DOI $:https://doi.org/10.1103/PhysRevE.100.022141$}
\bibitem{steadystate}
	\bibitem{steadystate}
Schnakenberg, Jürgen. "Network theory of microscopic and macroscopic behavior of master equation systems." Reviews of Modern physics 48.4 (1976): 571.
  \bibitem{seifert1}
  Barato, A. C., $\&$ Seifert, U. (2015). Thermodynamic uncertainty relation for biomolecular processes. Physical review letters, 114(15), 158101.
  \href{DOI:https://doi.org/10.1103/PhysRevLett.114.158101}{DOI: $https://doi.org/10.1103/PhysRevLett.114.158101$}
  \bibitem{seifert2}
  Seifert, U. (2018). Stochastic thermodynamics: From principles to the cost of precision. Physica A: Statistical Mechanics and its Applications, 504, 176-191.
  \href{https://doi.org/10.1016/j.physa.2017.10.024}{DOI: $https://doi.org/10.1016/j.physa.2017.10.024$ }
  \bibitem{currents_nature}
  Horowitz, J. M., $\&$ Gingrich, T. R. (2020). Thermodynamic uncertainty relations constrain non-equilibrium fluctuations. Nature Physics, 16(1), 15-20.
  \href{https://doi.org/10.1038/s41567-019-0702-6}{DOI : $https://doi.org/10.1038/s41567-019-0702-6$}
  \bibitem{kardar}
     Kardar, M. (2007). Statistical physics of particles. Cambridge University Press.
\end{thebibliography}
\end{document}